\documentclass[a4]{article}
\usepackage[utf8x]{inputenc}

\usepackage{color}
\usepackage{graphicx}
\usepackage{amsfonts}

\begin{document}
\newcommand{\RR}{\ensuremath{\textbf{R}}}
\newcommand{\UD}[2]{\ensuremath{^{#1}_{\phantom{#1}#2}}}
\newcommand{\UDD}[3]{\ensuremath{^{#1}_{\phantom{#1}#2 #3}}}
\newcommand{\UDDD}[4]{\ensuremath{^{#1}_{\phantom{#1}#2 #3 #4}}}
\newcommand{\DU}[2]{\ensuremath{_{#1}^{\phantom{#1}#2}}}
\newcommand{\KK}{\ensuremath{{Q}}}
\newcommand{\WT}[1]{\ensuremath{\widetilde{#1}}}
\newcommand{\MR}{\ensuremath{{\cal R}}}
\newcommand{\ML}{\ensuremath{{\cal L}}}
\newcommand{\WML}{\ensuremath{\widetilde{\cal L}}}
\newcommand{\HML}{\ensuremath{\widehat{\cal L}}}
\newcommand{\dd}{\ensuremath{\textrm{d}}}
\newcommand{\MK}{\ensuremath{{\cal K}}}
\newcommand{\tq}[1]{\ensuremath{ ^{#1}\!q}}

\newcommand{\blue}[1]{ {\color{blue} #1 } }

\def\p{\partial}
\def\lmax{l_{\textrm{\tiny max}}}
\def\rsc{\mathcal{R}}
\def\oq{ {^0\! q}}
\def\tx{ \hat{x}}
\def\tn{ \hat{n}}
\def\ll{\textrm{\tiny $\parallel$}}
\def\inv{\textrm{\tiny $(-1)$}}

 \title{Isometric embeddings of 2-spheres by embedding flow for applications in numerical relativity} 
\author{Michael Jasiulek\thanks{Max--Planck--Institut f\"ur Gravitationsphysik (Albert--Einstein--Institut), 
Am M\"uhlenberg 1, D-14476 Potsdam, Germany, E-mail: \emph{michaelj@aei.mpg.de} } \and Miko\l{}aj Korzy\'nski
\thanks{Gravitational Physics, Faculty of Physics, University of Vienna, A-1090 Vienna, Austria, E-mail: 
\emph{mikolaj.korzynski@univie.ac.at}}}

\maketitle

\begin{abstract}

We present a numerical method for solving Weyl's embedding problem which consists of finding a global isometric
embedding of a positively curved and positive-definite spherical 2-metric into the Euclidean three space.
The method is based on a construction introduced by Weingarten and was used in Nirenberg's proof of Weyl's
conjecture. The target embedding results as the endpoint of an embedding flow in $\RR^3$ beginning at the
unit sphere's embedding. We employ spectral methods to handle functions on the surface and to solve various
(non)-linear elliptic PDEs. Possible applications in $3+1$ numerical relativity range from quasi-local mass and
momentum measures to coarse-graining in inhomogeneous cosmological models.

\end{abstract}

\section{Introduction}

It is a classic result in differential geometry that for every 2-surface of $S^2$ topology, equipped with a
positive-definite metric whose curvature is positive, a global isometric embedding into the Euclidean three space
can be found. Moreover, the embedding is unique up to the isometries of the target Euclidean space which are
rigid motions: rotations and translations as well as reflections \cite{han-hong, spivak}. This was conjectured
by Hermann Weyl and later proven by Alexandrov and Pogorelov \cite{alexandrov, pogorelov}, and independently by
Nirenberg \cite{nirenberg-1953}.

Isometric embeddings into flat space have a wide range of applications in general relativity. For a given isometric
embedding in a curved ambient space they provide a reference surface, thereby fixing the intrinsic geometries.
From the difference between the extrinsic geometries it is possible to define measures of quasi-local mass like
the Brown-York and Kijowski-Liu-Yau masses \cite{kijowski-1997-29,Brown:1991gb,Brown:1992br,Liu:2003bx}. For more
examples see Szabados' review article \cite{szabados-2004-4}. Such quasi-local mass measures -- not yet applied
in (3+1) numerical relativity -- could potentially be useful to determine mass and momenta in binary black hole
simulations. In addition, the uniqueness of the reference shapes makes them ideal for visualising 2-metrics. 
Another application can be found in 3+1 numerical cosmology, where isometric embeddings of $S^2$ surfaces are the
principle step in a method for coarse--graining (averaging) of expansion and shear in inhomogeneous cosmological
models proposed by MK in \cite{Korzynski:2009db, Korzynski:2010np}.

The isometric embedding equations constitute a three-dimensional (3D) non-linear, coupled system of PDEs, for which
 standard numerical methods can not be applied and analytical solutions are impossible to find. This could be a
reason why no practical application in numerical simulations exists. The problem has already been addressed twice:
by Nollert and Herold in \cite{nollert}, and by Bondarescu, Alcubierre and Seidel \cite{bondarescu-2002-19}. In the
algorithm presented in \cite{nollert}, the surface in question is triangulated and then a corresponding polyhedron
in $\RR^3$ is constructed, whose edges have a length approximating those of the source triangulation. The method
has a serious drawback: multiple polyhedra exist fulfilling the length constraints between neighboring points, most
of them converge to unwanted non-regular (non-differentiable) surfaces. For the method in \cite{bondarescu-2002-19}
the three unknown embedding functions are expanded into harmonic polynomials of maximal degree $\lmax$ and the
embedding problem is turned into a numerical minimization problem of a $3 (\lmax+1)^2$ dimensional functional, which
vanishes only if the embedding is isometric. This renders the method crucially dependent upon the minimisation
algorithm used which, among other things, must steer away from the false minima of the functional. The minimization
becomes increasingly difficult and computationally costly by increasing the dimensionality of the parameter space. 

Our algorithm is based on the continuity method used in Nirenberg's proof and is not effected by the aforementioned
problems. Its convergence to the right solution is guaranteed by Nirenberg's theorem. The target embedding
results as the final surface of an embedding flow beginning with the unit sphere's embedding in $\RR^3$. In
each step of the flow the solution of the linearised embedding equation (LEE) allows one to continuously deform
the surface from a round unit sphere to the final shape. This requires the conformal factor linking target and
spherical round 2-metric, which we obtain as a steady-state solution of the Ricci flow. However we note that,
unlike the previous two approaches, it is strictly limited to surfaces whose curvature is positive everywhere:
Even if the isometric embedding exists for a surface with non-positive curvature, the algorithm can not be applied.

We present a variety of numerical methods, among others, a pseudo-spectral parabolic evolution scheme to solve the
various elliptic PDEs appearing in the problem that could be interesting for other purposes in numerical relativity. 

The paper is organized as follows: In the next section we provide the mathematical background for the paper, 
as well as description of the linearized embedding flow we use. In the third section we discuss the technical
details of the algorithm and its implementation; in the fourth, we present the results of a concrete numerical
test case. We state the final conclusions in the fifth section and in the appendix, where we also quote relevant
but more technical results obtained in this paper. 

\section{The mathematical background}

The Weyl problem can be described as follows: Given a 2-surface $C$ of spherical topology, equipped with a
positive-definite metric $q$ (target metric), find a sufficiently regular embedding into the Euclidean space
\begin{eqnarray}
 \Phi : C \to \RR^3
\end{eqnarray}
which preserves the 2-metric $q$, i.e.
\begin{eqnarray}
 q = \Phi^* \delta
\end{eqnarray}
where $\delta$ is the standard metric on $\RR^3$ and $\Phi^*$ denotes the standard pullback. In
this paper we shall assume that the metric and the embedding are smooth.

It is known that the embedding always exists if the metric is sufficiently regular and its curvature is positive
everywhere \cite{nirenberg-1953, han-hong}. It is also known to be unique up to the rotations, translation and
reflexions in the Euclidean space (see for example \cite{spivak} for a review of the rigidity results).

The algorithm to determine $\Phi$, which we present in this paper, is based on the continuity method used 
in Nirenberg's proof of existence and introduced by Weyl. It consists of three steps which we will now describe
briefly, leaving their detailed discussion until the next subsections.

At first we compute the conformal factor relating the target metric to the metric of the round sphere with radius
one $\oq$ (round metric) which is a steady-state solution of the Ricci flow.
\begin{equation}
 \oq = e^{- 2 \sigma }\, q 
\end{equation}
It is known that the Ricci flow uniformizes the metric \cite{hamilton-1988-71}, i.e. the flow converges to a constant
 curvature metric $q\rightarrow \oq$ whereby we obtain $\sigma$. In general, the round metric and $\sigma$ are
then given in arbitrary coordinates. This has to be corrected through a transformation, for which the coordinate
representation of $\oq$ and its embedding $\Phi_0$ into $\RR^3$ take a standard well-known form.

Then we construct a one-parameter family of metrics
\begin{equation}
 \tq{t} = \Omega(t,\sigma)^2\, \oq, 
\end{equation}
where $\Omega(t,\sigma)$ is a function chosen such that $\tq{t=1}$ is the target metric and $\tq{t=0}$
the round metric, which allows one to morph one metric smoothly into the other.

Finally, we perform the embedding flow: Beginning at $t=0$ the known standard embedding $\Phi_0$ of
the round metric is deformed in 'small' steps such that the induced metric of the deformed surface
at each step matches $\tq{t}$ until the target metric is reached. This is accomplished by solving
the linearized embedding equations, which relate a small change of the metric tensor with a small
deformation of the embedding functions. 

We now proceed to describe the three steps in detail. 

\subsection{The Ricci flow}

In the first step of our method we have to determine the conformal factor between the target and 
round sphere metric.
In Riemannian 2-manifolds the Ricci flow reduces to a single equation for the conformal factor 
\begin{eqnarray}
 p(\tau) &=& q\,e^{-2\rho(\tau)} \nonumber \\
 \dot \rho &=& \MR [ p(\tau) ] - \langle \MR [ p(\tau) ] \rangle,
\end{eqnarray}
where $\MR [p]$ is the Ricci scalar of the metric $p$ and $\langle \MR [ p ] \rangle$
its surface average wrt $q$ included to keep the surface area of $p$ bound.
$\MR [ p ]$ is related to the Ricci scalar of the target metric in the following way
\begin{eqnarray}
 \MR [ p ] = e^{2\rho}(\MR[q] + 2\, ^q\!\Delta \rho),
\end{eqnarray}
with $^q\!\Delta$ being the Laplacian wrt to $q$.

The flow converges for large $\tau$ to a function $\rho \buildrel {\tau \rightarrow \infty} \over \longrightarrow \sigma$
for which $e^{-2\sigma}\,q$ is a round sphere metric (see \cite{chow-1991-33, hamilton-1988-71},
the latter also in \cite{cao-book}). By simple rescaling of $\sigma$ we can ensure that the 
sphere has the area of $4\pi$. Now it is possible to construct a family of metrics joining target and 
round sphere metric 
\begin{eqnarray}
 \tq{t} = \,\tq{0}\,\,\Omega(t,\sigma)^2.
\end{eqnarray}
where $\Omega(t,\sigma)$ should be chosen such that $\MR [ \tq{t} ]$ is always positive. This is 
guaranteed for $\Omega(t,\sigma) = e^{2\sigma \, t} $. We have also tested a different function to drive
the embedding flow $\Omega(t,\sigma) = t(e^\sigma - 1) + 1 $, in which the change between consecutive embeddings in
the flow, as explained later, is linear in $t$ as well. 

\subsection{Round metric in standard coordinates}

The steady-state solution of the Ricci flow is a round metric $\tq{0} =e^{-2\sigma}\,q$
in arbitrary coordinates.
To proceed, a transformation to standard coordinates $x^i$ is required for which its isometric
embedding $X^i$ into $\RR^3$ takes the well-known form, unique up to rotations and reflections
\begin{eqnarray} 
 X^1 &= x/r =&\sin\theta\,\cos\varphi\nonumber\\
 X^2 &= y/r =&\sin\theta\,\sin\varphi\nonumber\\
 X^3 &= z/r =&\cos\theta, \label{eqRoundEmbedding}
\end{eqnarray}
where $r=\sqrt{x^2+y^2+z^2}$ and $(\theta,\varphi)$ are spherical coordinates corresponding to $x^i$. Note that
the three functions on the right hand side of (\ref{eqRoundEmbedding}) constitute a real, orthogonal basis for
the spherical harmonics with $l=1$. This means that these functions, denoted as $n^i$, are eigenfunctions of the
Laplace operator on the sphere with eigenvalue $-2$
\begin{eqnarray}
 \Delta n^i = -2 n^i,\quad i=1,2,3 \label{eqLaplaceEigen}.
\end{eqnarray}
They are also orthogonal and normalized in the sense that
\begin{eqnarray}
 \oint_C n_i\,n_j\,\dd A = \frac{4\pi}{3}\,\delta_{ij}, \label{eqfNormalization}
\end{eqnarray}
where $\dd A$ is the area element associated with the round metric. It is easy to check that the opposite is
also true: Any three orthogonal and properly normalized functions satisfying (\ref{eqLaplaceEigen}) are related
to (\ref{eqRoundEmbedding}) by an $O(3)$ linear mapping and therefore constitute an isometric embedding of the
unit sphere themselves. Equation (\ref{eqLaplaceEigen}) in turn can be solved numerically even in non-standard
coordinates.

\subsection{The linearized embedding equation}
\label{subslinemb}
Consider an $S^2$ surface $D$, endowed with a coordinate system $\theta^A$ and embedded in $\RR^3$ by a mapping 
described by three functions $X^i(\theta^A)$. The induced metric has the form of
\begin{eqnarray}
 q_{AB} &=& X\UD{i}{,A}\,X\UD{j}{,B}\,\delta_{ij}.
\end{eqnarray}
If we deform the embedding by adding a small $\delta X^i(\theta^A)$, the metric changes according to
\begin{eqnarray}
 \delta q_{AB} &=& 2\,\delta X\UD{i}{\left(,A\right.}\,X\UD{j}{\left.,B\right)}\,\delta_{ij} \label{eqLEE}
\end{eqnarray}
up to the linear order. Given the metric change $\delta q$ one can ask for the compatible deformation vector.
Finding $\delta X^i$ involves solving (\ref{eqLEE}), which is called the linearized embedding equations (LEE).
Through a variable transformation this linear system of three PDEs can be turned into to a single elliptic equation
of the second order for Weingarten's ``Verschiebungsfunktion'' and two ODEs, see also \cite{han-hong,nirenberg-1953}
for derivations. Let $Y^i$ denote the deformation vector field we seek and $d_{ij}$ the metric deformation and
let $s^i$ be the outward-pointing null normal. We decompose $Y^i$ into the normal and tangential part:
\begin{eqnarray}
 Y^i &=& \Upsilon\,s^i + I^i \nonumber \\
\end{eqnarray}
and introduce new variables $u_A$ and $w$
\begin{eqnarray}
 u_{A} &=& s_k\,Y\UD{k}{,A} \nonumber \\
 w &=& -\epsilon^{AB}\,D_{A} \nonumber I_B. 
\end{eqnarray}
$\epsilon^{AB}$ denotes the area form and $D_A$ is the covariant derivative on the surface.
$Y^i$ can be reconstructed from $u_A$ and $w$ via
\begin{eqnarray}
 \Upsilon_{,A} &=& u_A - K_{AB}\,I^B \nonumber \\
 D_A I_B &=& -\frac{1}{2} w\,\epsilon_{AB} + \frac{1}{2} d_{AB} + \Upsilon\,K_{AB} \label{eqDI},
\end{eqnarray}
where $K_{AB}$ denotes the extrinsic curvature. On a surface of positive scalar curvature $K_{AB}$ is positive
definite \footnote{This is the precise reason why our method is limited to positively curved metrics.} and therefore
has an inverse $(K^{-1})^{AB}$. The variable $u_A$ is then related to the first derivative
of $w$ by
\begin{eqnarray}
 u_A &=& \frac{1}{2}\epsilon_{AB}\,\left(K^{-1}\right)^{BD}\left(c_{D} - w_{,D}\right), \label{equA} \\
 c_D &=& -D_{A}q_{BD}\,\epsilon^{AB} \nonumber.
\end{eqnarray}
Finally $w$ itself has to satisfy an elliptic equation
\begin{eqnarray}
 \ML w &=& \tau \label{eqLw}.
\end{eqnarray}
The elliptic operator $\ML$ is defined as
\begin{eqnarray}
 \ML w &=& D_A\,\left((K^{-1})^{AB}\,w_{,B}\right) + \MK\,w, \label{eqLdef}
\end{eqnarray}
where $\MK = K\UD{A}{A}$, and the right hand side of the equation is
\begin{eqnarray}
 \tau &=& D_A\,\left((K^{-1})^{AB}\,c_{B}\right) - K\UD{C}{A}\, d_{BC}\,\epsilon^{AB}. \label{eqtaudef}
\end{eqnarray}
Equation (\ref{eqLw}), together with (\ref{equA}) and (\ref{eqDI}), is equivalent to the original LEE.

$\ML$ is self-adjoint with the standard scalar product $\langle f,g\rangle = \int_C f^*\,g\,\dd A[q(t)]$
and thus has only real eigenvalues.
Moreover, for any convex surface it has a three-dimensional kernel spanned by the components of the
normal vector $s^1(\theta^A)$, $s^2(\theta^A)$ and $s^3(\theta^A)$ \cite{han-hong}. In case of a
round sphere it is possible to demonstrate that $\ML= \Delta + 2$.

Since $\ML$ has a non-trivial kernel, it is not invertible. Nevertheless, equation (\ref{eqLw}) has solutions if
its right hand side is orthogonal to the kernel:
\begin{eqnarray}
 \left\langle \tau, s^i\right\rangle = 0.
\end{eqnarray}
The solution is unique up to adding a combination of the functions $s^i,\, i=1,2,3$. Geometrically this ambiguity
corresponds to the possibility of adding a rigid rotation generator to
solutions of (\ref{eqLEE}).

\section{Technical and numerical details}%

In this section we explain the numerical and technical details of our implementation. Our approach consists of
three main steps: the Ricci flow to relate target and round metric, the $l=1$ eigenvalue problem of the round
metric Laplacian in non-standard coordinates, and the embedding flow from the round metric's embedding to the target
embedding. For each computational step we need to conduct a variety of high accuracy numerical operations on spherical
surfaces (numerical integration, interpolation, (anti)-differentiation, coordinate inversion, solving elliptic
PDEs). For this reason spectral methods in combination with particular coordinates are the best choice as we
explain in the following.

\subsection{Coordinate basis, polynomial basis, grid setup}%

Instead of covering spherical surfaces with 2D coordinate maps we consider 2-surfaces as being embedded in some
(fictitious) ambient Riemanian 3-space (for example the Euclidian space), equipped with some 3D quasi-Cartesian
coordinate system $\{ \hat{x}^i \}$ and represent all surface tensors using this exterior coordinate basis.
Polar coordinates $(\hat{\theta},\hat{\phi})$ on the surface are merely used to label the grid points
$(\hat{\theta}_i, \hat{\phi}_j),\, i=1,\cdots,N_{\hat{\theta}},\,j=1,\cdots,2 N_{\hat{\theta}}$.
We use a Gauss-Legendre grid structure, the \emph{canonical} grid, on which a surface integral of polynomials
of degree $2 \lmax=2 (N_{\hat{\theta}}-1)$ can be represented exactly by a finite sum.

Our approach requires various numerical operations on the surface: function evaluations at non-canonical points (``eval''),
numerical integration, function inversion (``inv''), differentiation and anti-differentiation (``int''). 
For this reason we chose to represented the shape function $h$ and other surface tensors as an expansion in harmonic polynomials. 

\begin{equation}
 h = \sum^{l_{max} }_{lm} {^\Phi} [h]^{lm} \, \Phi^{lm} + \mathcal{O}(\lmax+1) = 
  \sum^{l_{max} }_{lm} {^Y} [h]^{lm} \, Y^{lm} + \mathcal{O}(\lmax+1), 
\end{equation}
where $Y^{lm}$ is the standard othonormalised basis and the other basis is defined by $\Phi^{lm} := (\hat{n}^i \, \mathcal{N}^{lm}_i)^l$,
where $\hat{n}^i = \hat{x}^i/\hat{r} = ( \sin\hat{\theta}\,\cos\hat{\phi}, \sin\hat{\theta}\,\sin\hat{\phi}, \cos\hat{\theta} )$
is the radial unit normal. This basis is orthogonal wrt distinct $l$-eigen-spaces. $\mathcal{N}_i$ is a list of constant complex
null vectors that span the $2l+1$ harmonics in each eigenspace. The null vectors are chosen as in \cite{Jasiulek:2009zf}, then both basis are
related by a discrete Fourier transform in each eigenspace.

\subsection{Differentiation on the surface}%

The simple form of the basis $(n^i \, \mathcal{N}^{lm}_i)^l$ is practical for evaluation of functions off the grid. In addition its
differentiation\footnotemark is straightforward, for example the derivative in 3d coordinates takes the form of
\footnotetext{Alternatively, the differentiation of an expansion in $Y^{lm}$s can be performed by evaluating $\p_i,\p_{ij} Y^{lm}$ from $\p_i,\p_{ij}\Phi^{lm}$. 
It is practical to tabulated the $\p_i,\p_{ij} Y^{lm}$ initially.}
\begin{equation} \label{eq:partial_h}
 \partial_i h = \partial_i n^k \, \sum_{lm} {^\Phi} [h]^{lm} (n^j \, \mathcal{N}^{lm}_j)^{l-1} l \,\mathcal{N}^{lm}_k.
\end{equation}
By placing $ \p_i n^k \rightarrow \p_\theta n^k$ in front of the sum above, one obtains $\p_\theta h$ etc.; another practical feature.

\begin{figure}[t]
\includegraphics[width=0.49\linewidth,bb=134 380 476 616,clip]{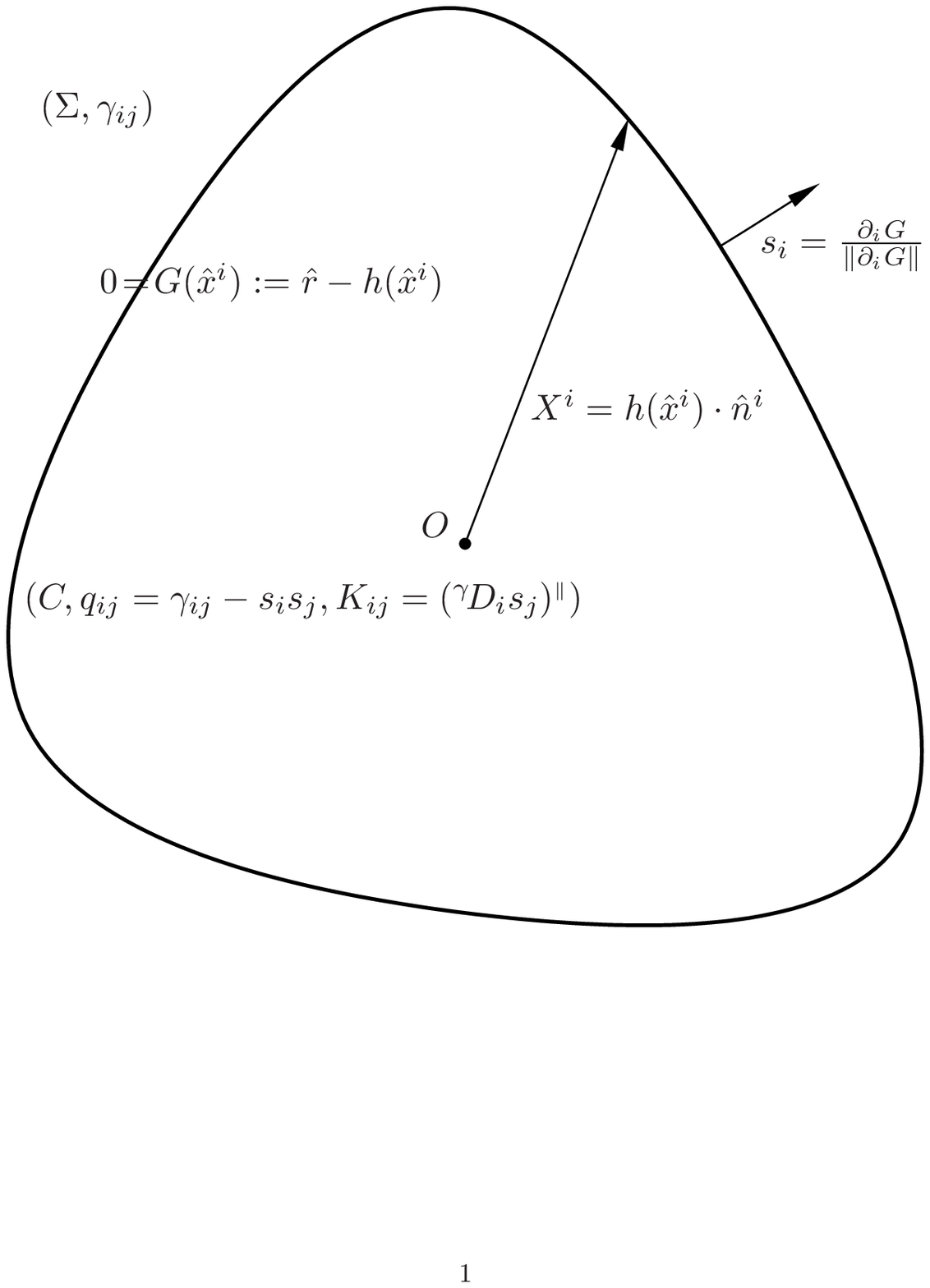} \quad
\includegraphics[width=0.49\linewidth,bb=133 342 473 545,clip]{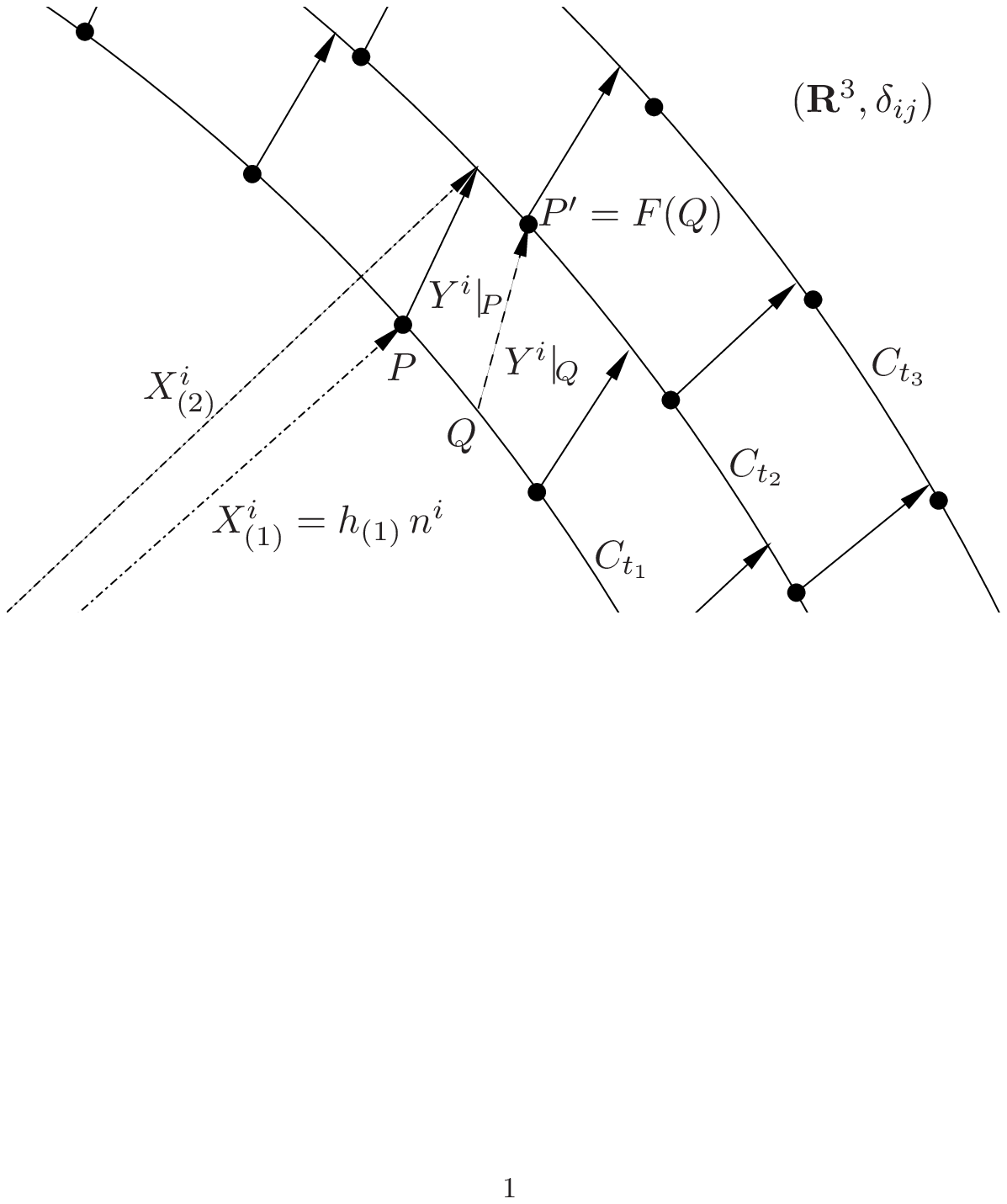}
\caption{ Left: Level surface $(C,q_{ij})$ given by $G(\hat{x}^i)= \hat{r} - h(\hat{x}^i)$ at $G=0$ and induced metric $q_{ij}$
  in the Riemanian manifold $(\Sigma,\gamma_{ij})$.
  Right: Embedding flow at $t=t_1,t_2,t_3$ in the Euclidean space. Shift vector field $Y^i$ at each step
   induces a new embedding $F: X^i_{(1)} \mapsto X^i_{(2)}=X^i_{(1)}+ Y^i$ that is in general off the
   canonical grid (dots). 
\label{fig:emb_flow}}
\end{figure}

We prescribe 2-surfaces through level set functions $G$ / shape functions $h$, see fig. (\ref{fig:emb_flow}) (left),
in some ambient manifold $(\Sigma,\gamma_{ij})$, as is common in numerical simulations of the Einstein equation in (3+1) dimensions.
If a shape function $h$ is given, the computational steps to calculate curvature tensors on the surface are as follows
\begin{equation} \label{eq:h-steps}
\begin{tabular}{ccccccccc}
 $h$ & $\rightarrow$ & $[h]^{lm}$ & $\rightarrow$ & $\p_i h,\, \p_{ij} h$ & $\rightarrow$ & $s_i, q_{ij}, K_{ij}$ & $\rightarrow$ & $\mathcal{R}, \blue{K^\inv_{ij} }$ \\
 &  &  &  & $\blue{\p_{ijk} h}$ & $\rightarrow$ & $\blue{\p_i K_{jk}}$ & $\rightarrow$ & $\blue{ \p_i K^\inv_{jk} }$, 
\end{tabular}
\end{equation}
where the expressions in blue are only necessary for solving the LEE. 

The computation of curvature tensors on the surface requires $^q\! D_i \chi_j$ of tensors $\chi_i$, which 
might have a normal component $\chi_s \neq 0$, as for example $\p_i h$, 
\begin{equation}
 {^q\!D}_i \chi_j = (^\gamma\! D_i \chi_j)^\ll - K_{ij} \chi_s,
\end{equation}
which is not simply the tangential part of the ambient derivative $(^\gamma\! D_i \chi_j)^\ll$.
As a consequence, the Laplacian of a function $\psi$ on $C$ is calculated by
\begin{equation} \label{eq:laplace}
 {^q\! \Delta} \psi = q^{ij} \p_{ij} \psi - {^\gamma \Gamma}^i \p_i \psi - \mathcal{K}\, \p_s \psi,
\end{equation}
where ${^\gamma \Gamma}^i = {^\gamma \Gamma}^i_{jk}\, q^{jk}$ is this contraction with the Christoffel symbols of $\gamma_{ij}$. 

\subsection{Parabolic flow relaxation method}%

The elliptic PDEs of type $\mathcal{L}(u) - V(u)=0$ appearing in our approach are solved by considering the 
associated parabolic flow equation \footnotemark
\footnotetext{ In principle, the spectral decomposition of functions on the surface would allow us to turn non-linear / linear elliptic PDEs into
algebraic / linear systems of equations to be solved with Newton-Raphson's / splitting (relaxation) methods which are numerically more efficient but
more stringent with regard to the initial guess and the conditioning of involved matrices. The parabolic flow relaxation method on the other hand 
covers non-linear / linear elliptic PDEs and is more robust. Moreover, efficiency is not a concern for the 2D PDEs we deal with.}
\begin{equation} \label{eq:para_pde}
 \dot{u} = \mathcal{L}(u) - V(u) \quad \buildrel {- \langle \textrm{rhs} \rangle} \over \longrightarrow \quad 
 \dot{u} = \left\{\mathcal{L}(u) - V(u) \right\}^{/l=0} 
\end{equation}
where $\dot{u}$ is the time derivative of $u$ and $\left\{ \bullet \right\}^{/l=0} := \bullet - \langle \bullet \rangle$
removes a function's surface average ($l=0$ mode). 
For linear elliptic operators $\mathcal{L}$ with a non-positive spectrum, except $l=0$,
arbitrary initial data evolves to a steady-state solution (at $\dot{u},\, \ddot{u},\, \cdots =0$) of the 
parabolic PDE which is automatically a solution to the elliptic equation. The Ricci flow
is a parabolic PDE with a non-linear elliptic part. The existence of its steady states and convergence to them is 
known \cite{chow-1991-33, hamilton-1988-71}. 

The parabolic PDEs we deal with are solved by means of a pseudo-spectral scheme, where we use the method of lines, 
2. or 4. order Runge-Kutta (RK2, RK4) and spectral finite difference methods, see eq. (\ref{eq:partial_h}), until $| \dot{u} |$ 
falls below machine/spectral precision. Since we are not interested in the details of the solution during 
the relaxation period, but in a quick fall-off, we pick a coarse time-step $\Delta t$ and RK2 (Heun), where we take the CFL-condition
of the heat equation with explicit Euler $\Delta t < \nu (\Delta x)^2,\,\nu=0.25$ as an orientation.

The algebraic operations to compute the rhs produce an aliasing error. Therefore, we filter modes $> \lmax$ form the rhs. 

\subsection{Computational steps: Ricci flow, EVP, LEE} \label{subsec:comp-steps}

The computational steps of the main component 'Ricci flow' of our implementation are the following. 
A shape function $h(\tx)$ and an ambient geometry, see fig. (\ref{fig:emb_flow}) (left), are required as input
\begin{center}
\begin{tabular}{ccccccc}
 $h(\tx)$ & $\rightarrow$ & $ \textrm{eq.}\,(\ref{eq:h-steps}) $ & $\rightarrow$ &
 $\textrm{evolve:}\,\, \dot{\sigma} = \{ \rsc[e^{2\sigma} q ] \}^{/l=0}$ & $\rightarrow$ & $\sigma(\tx), \oq(\tx)_{ij}$,
\end{tabular}
\end{center}
where $\rsc[e^{2\sigma} q ] = e^{2 \sigma} \left( \rsc[q] + 2\, ^q\!\Delta\sigma \right)$ and $^q\!\Delta\sigma$ is computed as in eq. (\ref{eq:laplace}).
The spherical round metric $\oq_{ij} = e^{-2 \sigma} q_{ij}$ with $\rsc[\oq] \equiv 2$ is the steady-state solution of the Ricci flow. 
It is given in non-standard coordinates in which its Laplacian takes the form $^0 \Delta \bullet = e^{2 \sigma}\, ^q\! \Delta \bullet$.
As a standard coordinate basis, we take the three orthonormalised $l=1$ eigenfunctions $n^j$ of $^0 \Delta$, see eq. (\ref{eqRoundEmbedding}) 

\begin{center}
 $\sigma(\tx)$ $\,\,\rightarrow\,\,$ $\textrm{evolve:}\,\, \dot{n}^i = \{ ^0\! \Delta n^i + 2 n^i \}^{/l=0}$ $\,\,\rightarrow\,\,$ $n^j(\tx)$ $\,\, \buildrel \textrm{inv} \over \longrightarrow\,\,$ $\tn^i(x)$
 $\,\,\buildrel \textrm{eval} \over \rightarrow\,\,$ $\sigma(x)$
\end{center}
in which $\oq$ takes the well known simple form $\oq_{ij}=\delta_{ij} - n_i n_j$. Identifying $x^i$ with
Cartesian coordinates in $\RR^3$ maps $\oq_{ij},\sigma$ to the Euclidean space where the embedding flow is performed,
see fig. (\ref{fig:emb_flow}) (right).
We construct a family of metrics $\tq{t}_{ij} = \Omega(t,\sigma)^2\cdot \oq_{ij}$ 
such that $\tq{t=0}\equiv\oq$ and $ \tq{t=1} \equiv q$ with $\rsc[ \tq{t} ] > 0$ and divide the time interval into $N$ steps,
i.e. $t=0,t_1,t_2,\cdots,t_N$ \footnotemark whereby we slowly deform $\tq{t}$ into the target metric so that the difference
\footnotetext{The number of steps and the values $t_n$ are arbitrary and depend on how far $q$ is off the round metric, usually $3<N<7$.}
\begin{equation} \label{eq:eps_metrics}
 ^{(n)} d_{ij} := q(t_{n+1})_{ij} -\, ^{(n)}\! q_{ij} 
\end{equation}
is small, where $^{(n)} \! q_{ij}$ is the induced metric wrt $h_{(n)}$, see fig. (\ref{fig:emb_flow}) (left), at the $n$th step. 
We initialize the flow with the shape function $h_{(0)} = 1$ corresponding to the standard embedding of the unit sphere.
At every step we compute
\begin{center}
 $h_{(n)}$ $\,\, \rightarrow \,\,$ eq. (\ref{eq:h-steps}) $\,\, \rightarrow \,\,$ compute: $^{(n)} d_{ij}$ $\,\, \rightarrow \,\,$
 evolve: $\dot{w} = \{ ( \mathcal{L}(w) - \tau)/ \mathcal{K} \}^{*}$ \\
\vspace{0.1cm}
 $\,\,\rightarrow\,\,$ $w$ $\,\, \buildrel { \textrm{int} } \over \rightarrow\,\,$ 
 $Y^i\vert_P$ $\,\,\rightarrow\,\,$ $n^i_{(n+1)}( x\vert_P)$ $\,\, \buildrel { \textrm{inv} } \over \rightarrow\,\,$ $n^i( x_{(n+1)}\vert_{P'})$ $\,\, \buildrel { \textrm{eval} } \over \rightarrow\,\,$ 
 $(Y^i, \p_jY^i, \sigma, q_{ij} )\vert_Q $ \\
\vspace{0.1cm}
 $\,\,\rightarrow\,\,$ $(h_{(n+1)}, q_{ij})\vert_{P'}$,
 
\end{center}
where $*$ denotes projecting out the lowest eigenvalue of $\ML$ (see Appendix).
The complications arise from the fact that the new embedding induced by $Y^i$ is shifting the points off the canonical grid, moreover, 
the target metric ($\oq$ and $\sigma$) has to be transported under the mapping $ X^i_{(n+1)} = X^i_{(n)} + Y^i$ from one surface to the other. 
\begin{equation} 
 q_{ij}\vert_{P'} = \frac{\p X^k_{(n)} }{\p X^i_{(n+1)} } \frac{\p X^l_{(n)} }{ \p X^j_{(n+1)} } \, q_{kl}\vert_Q
\end{equation}
After $N$ equi-distant steps $h_{(N)}$ the difference (\ref{eq:eps_metrics}) between the target metric and the induced metric wrt 
$h_{(N)}$ is of the order $N\cdot\mathcal{O}( d )$.
But this solution can be refined by recursively applying \footnotemark the above computational steps with $\tq{t=1}_{ij}$ fixed
in eq. (\ref{eq:eps_metrics}) whereby $d_{ij}$ converges to zero exponentially. 
\footnotetext{If necessary this refinement process can be applied at any intermediate time $t_n$, such that the embedding flow does not move to far off the conformal metric flow.}

\section{Numerical test case}

\begin{figure}[t]
\includegraphics[width=0.49\linewidth]{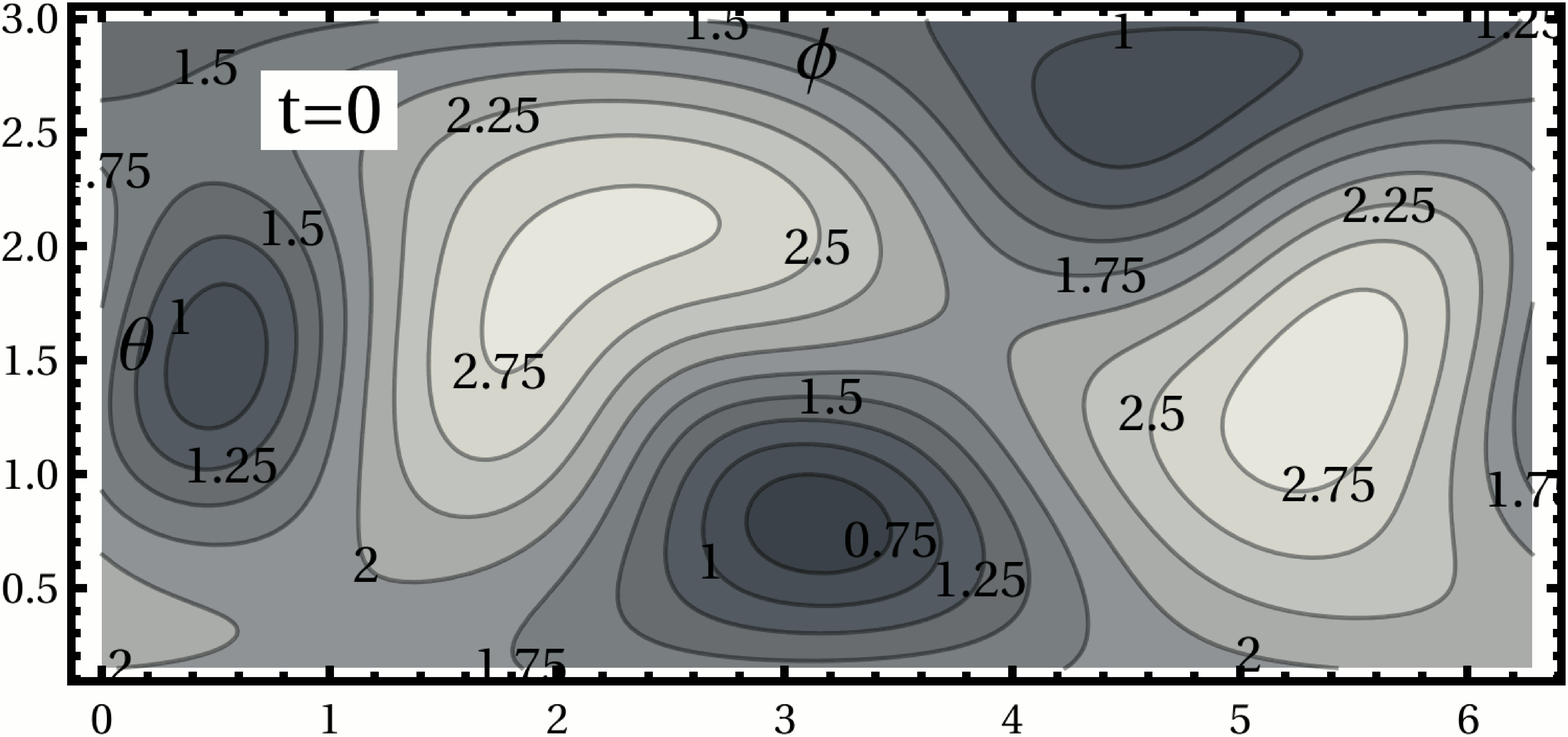} \quad
\includegraphics[width=0.49\linewidth]{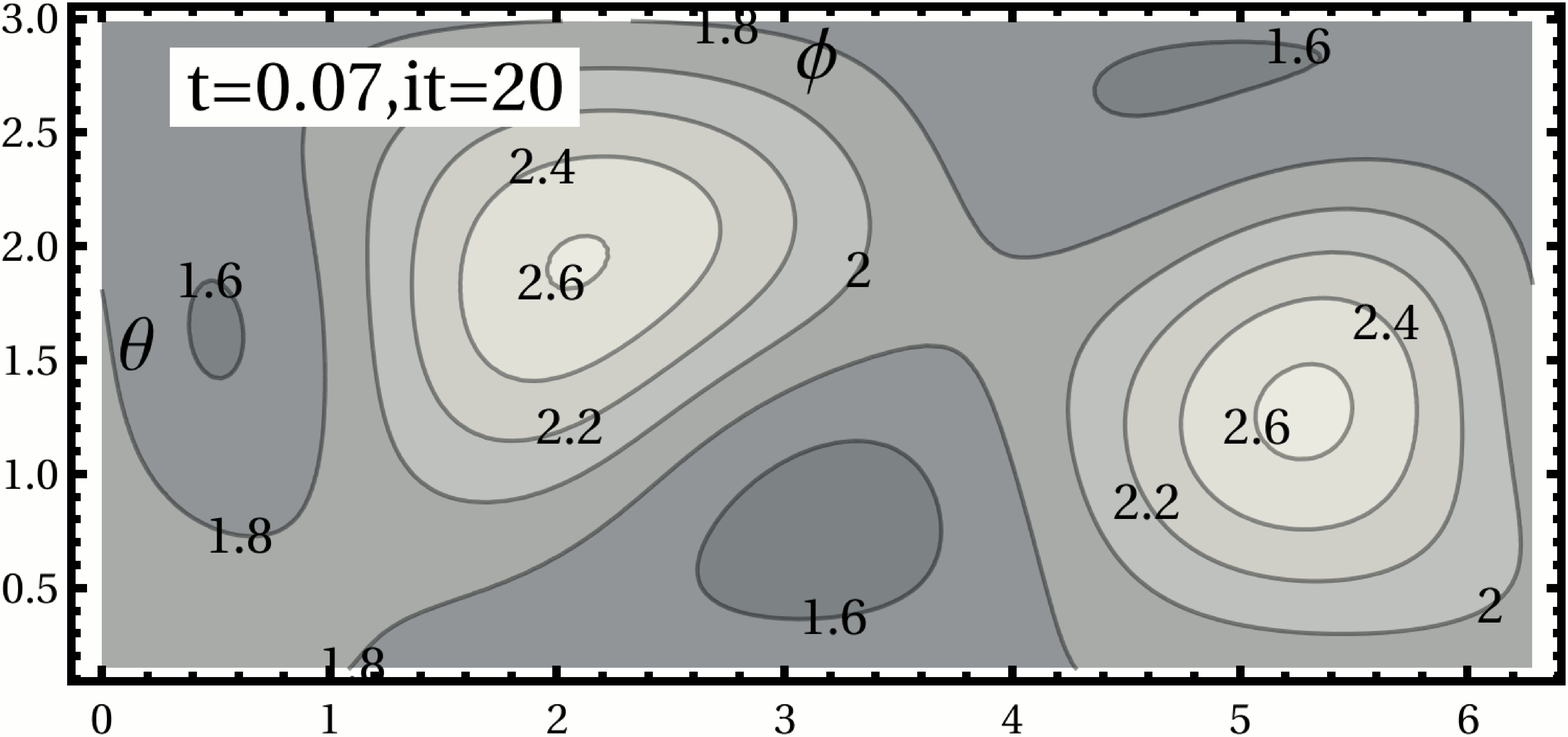}
\caption{ $(\theta,\phi)$-contour plots of the scalar curvature $\rsc[\tq{t}]$ in relaxation flow at $t=0$ and $t=0.07$. 
\label{fig:ricci_flow}}
\end{figure}

In this section we apply our procedure to a numerical test case, where we prescribe a test shape function in the
Euclidean space given by the function below. We solve the embedding problem numerically and compare the original
embedding with the resulting one. More specifically, we compute the induced 2-metric, perform the Ricci flow to
the round sphere, solve the EV problem and perform the embedding flow from the round sphere to the target shape,
solving the linearized embedding equation at each step.

As a test case we pick a cigar-shaped function (for illustration purposes) and add a randomized non-polynomial part 
\begin{eqnarray}
 h = c_1 \left( 1 + c_2 Re[ (\mathcal{N}^j_1 n_j) ]^2 + \sqrt{ 1 + c_3 Re[ (\mathcal{N}^j_2 n_j)^2 + (\mathcal{N}^j_3 n_j)^3 ] } \right) \nonumber
\end{eqnarray}
$c_{1,2,3}=0.458,\, 0.5,\, 0.15$ (for which the area $A\approx 1$ and $\rsc>0$, $\mathcal{K}>0$) and $\mathcal{N}_{1,2,3}$ \footnotemark 
are three randomly oriented null vectors. This way all $lm$ modes are occupied in a somewhat random manner. 
The surface defined by this shape function has the following area and central $n$-moments of the curvature $^n\sqrt{\mu_n(\bar{\rsc})}$:
\footnotetext{ $Re[ \mathcal{N}_{1,2,3} ] = 1/\sqrt{434}(-9,17,-8),$ $1/\sqrt{3786} (-1,-43,44),$ $1/\sqrt{78}( 7,-2,-5)$ 
and $Im[ \mathcal{N}_{1,2,3} ] = -1/\sqrt{3} (1,1,1). $ }
\begin{eqnarray}
 A/4\pi = 0.999285,\,\quad&& ^2\sqrt{\mu_2(\bar{\rsc})} = 0.280044,\nonumber\\
 \,^3\sqrt{\mu_3(\bar{\rsc})} = 0.197685,\,\quad&&^4\sqrt{\mu_4(\bar{\rsc})} = 0.342832, \nonumber
\end{eqnarray}
\begin{figure}[t]
\includegraphics[width=0.4\linewidth]{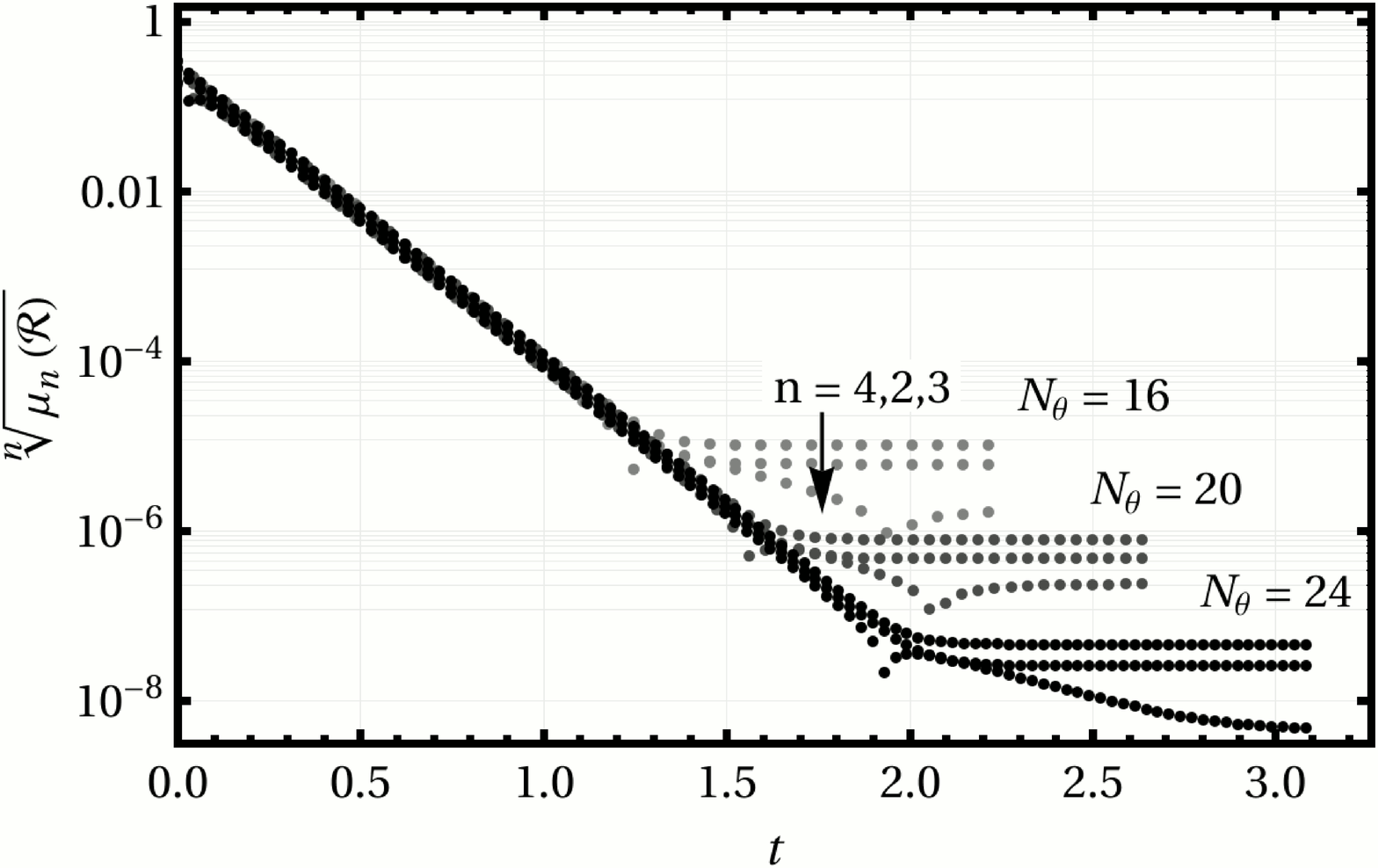} \quad
\includegraphics[width=0.53\linewidth]{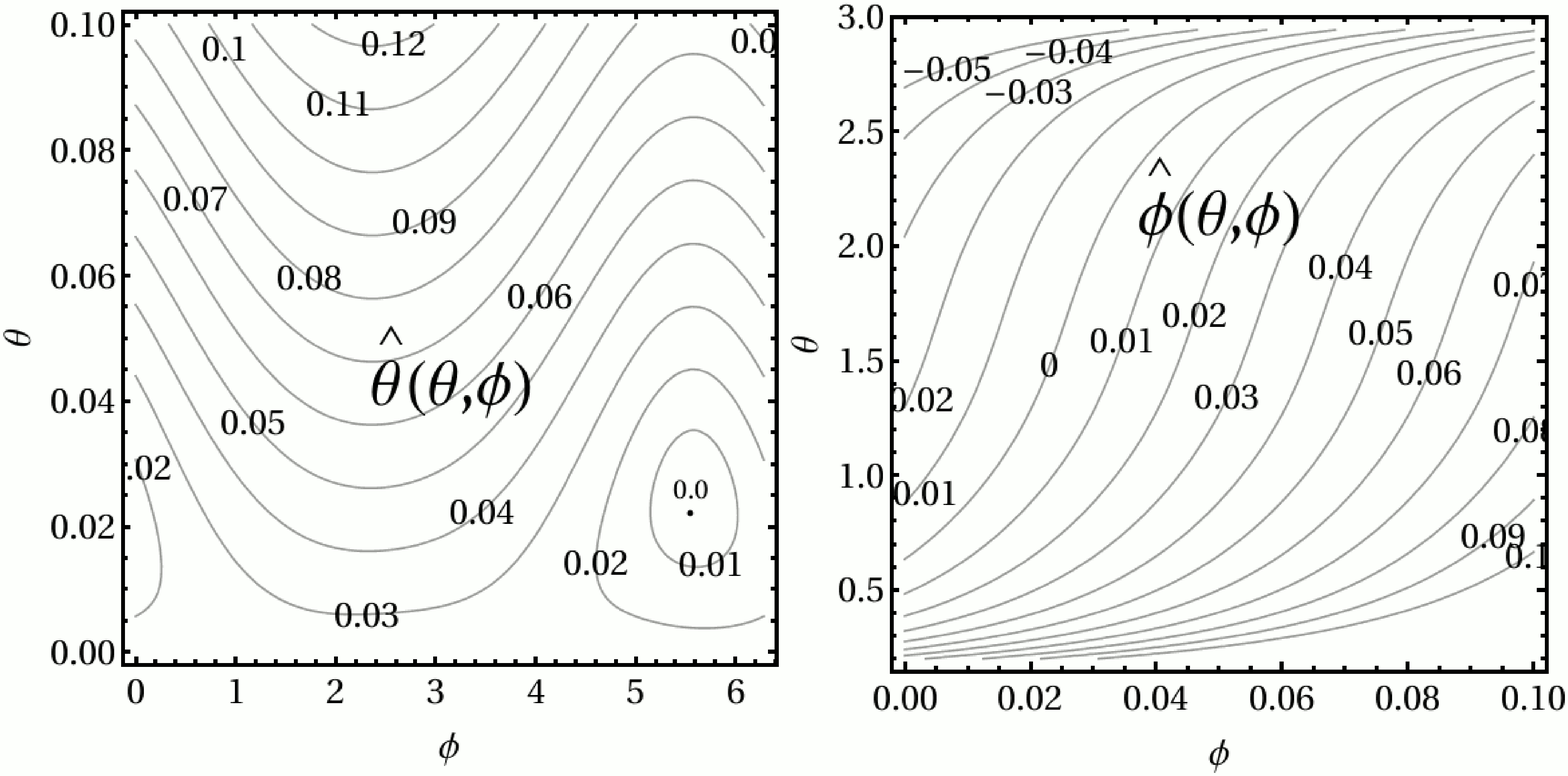} 
\caption{Left: Exponential fall-off of $^n\sqrt{ \mu_n( \bar{\rsc})}$ for $n=2,3,4$ at three different resolutions $N_\theta=16,20,24$.
  Right: ``old'' $(\hat{\theta}, \hat{\phi})$-coordinates given in the new $(\theta,\phi)$-coordinates. 
\label{fig:moments-coords}}
\end{figure}
where
\begin{eqnarray}
 \mu_n(\bullet) := \langle ( \langle \bullet \rangle - \bullet )^n \rangle, \nonumber
\end{eqnarray}
and $\bar{\rsc}=c\, \rsc$ with $\langle \bar{\rsc} \rangle = 1$. 
As shown in fig. (\ref{fig:ricci_flow}) (left) the scalar curvature significantly differs from that of a
round sphere initially, where $\sigma=0$ is set, but smoothes out towards $\rsc\approx 2$ in the Ricci flow
(right), where we employ the relaxation method as explained in the last section. The fall-off is exponential,
see fig.(\ref{fig:moments-coords}) (left), until it reaches a plateau of truncation error which converges to
zero by increasing the spherical resolution. The second order scheme is computationally more efficient, since
the parabolic flow is stable up to $\nu=0.11$ for RK2-Heun and up to $\nu=0.16$ for RK4 \footnotemark (measured
for $N_\theta=16$, $dx = 0.186$). \footnotetext{ On the same surface solving the heat equation we got $\nu=0.20$
RH2-Heun, $\nu=0.28$ RH4.}

During the Ricci flow the surface area ( $A/(4\pi) = 0.999285$ ) is bound but not fixed through
eq.(\ref{eq:para_pde}). The resulting round sphere metric $e^{-2\sigma} q_{ij}$ has the total area of $A/(4\pi)
= 1.007889$, which we normalise to unity. 
In the next step, we compute the $l=1$ eigenfunctions $n^j$ of this metric, as explained in
Sec. \ref{subsec:comp-steps}. As initial data for the relaxation method we picked $n^j|_{t=0}=\hat{n}^j$, where the
parabolic flow was stable up to $\nu=0.23$ for RK2. The three steady-state solutions are orthonormalised through
the Gram-Schmidt process and inverted $n^j( \hat{n}) \rightarrow \hat{n}^j( n )$ using Newton-Raphson's method
(since we have access to $\p_i n^j$ ). In fig. (\ref{fig:moments-coords}) (right) the old spherical coordinate
lines $\hat{n}^j(n)$ are shown on the new grid on which we re-evaluate $\sigma$ / the spherical round metric $^0
q_{ij}$. The round metric in the canonical coordinates $n^i$ can be embedded into Euclidean space by the shape
function $h_{0}=1$. During the embedding flow the shape is stepwise 'deformed' into the target shape function by
solving the LEE at each step. Again the relaxation method with RK2 for $\nu=0.2$ is employed.

\begin{figure}[t]
\begin{center}
\includegraphics[width=0.53\linewidth]{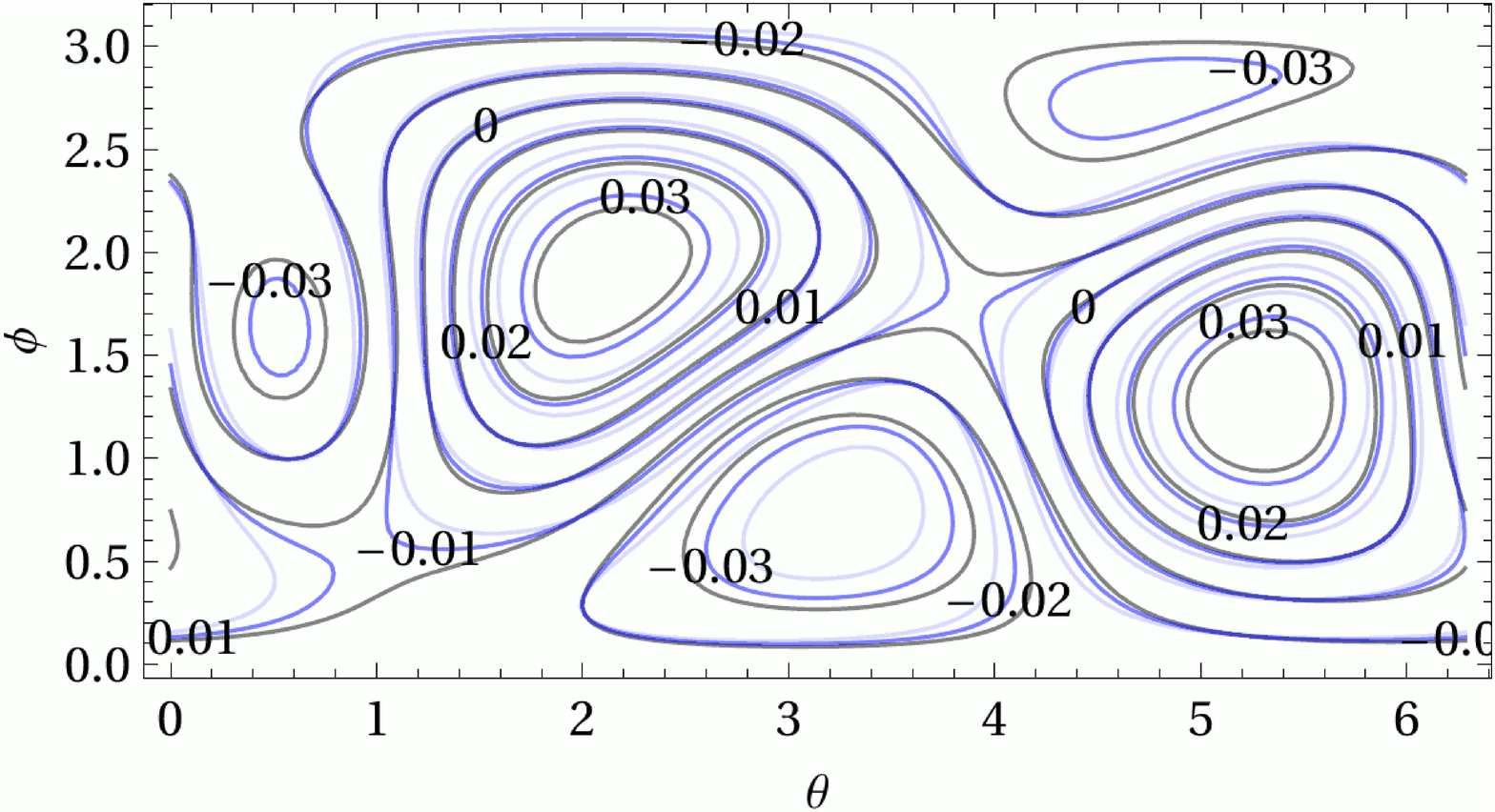} 
\includegraphics[width=0.45\linewidth]{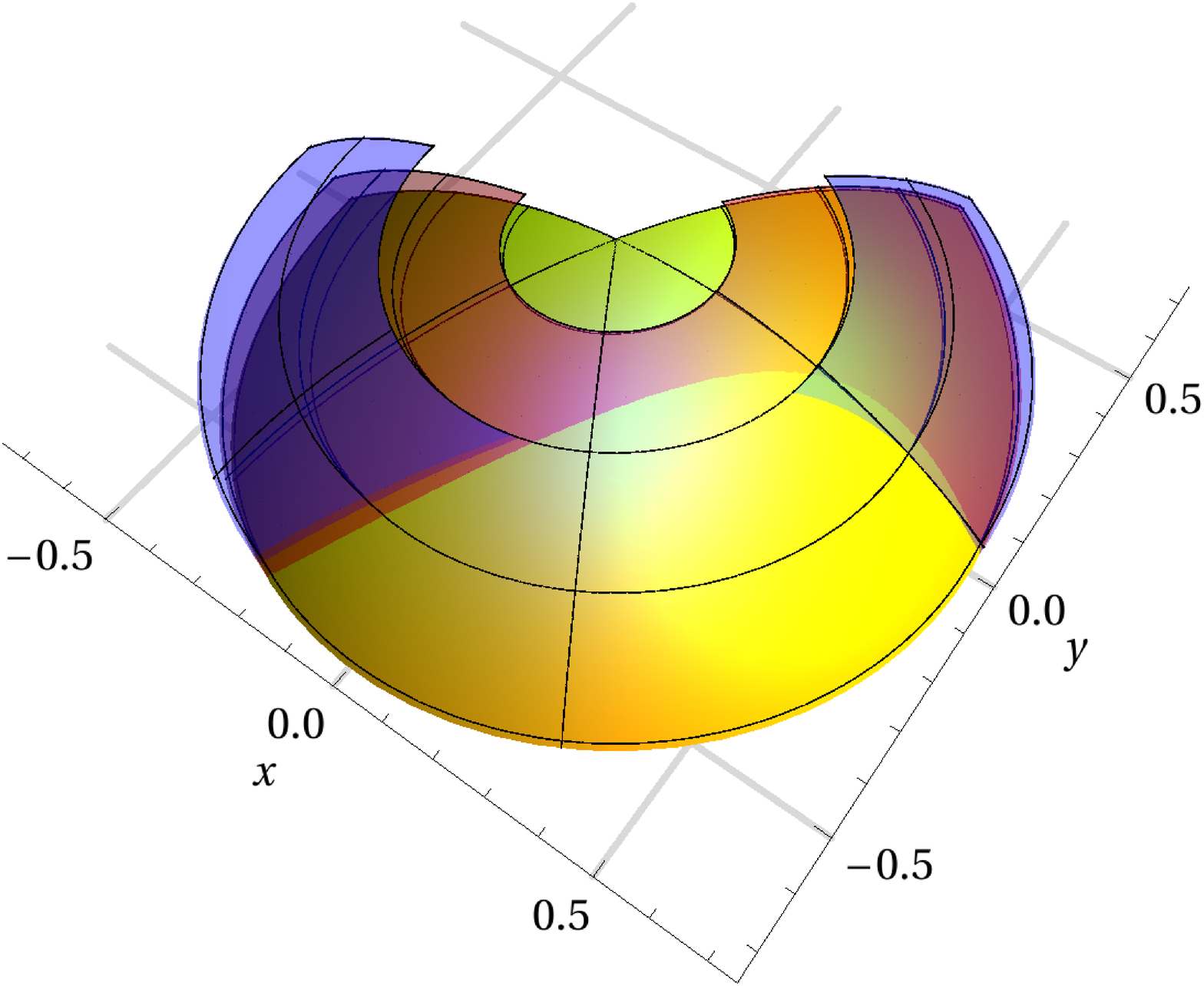}
\end{center}
\caption{ Left: Contours of difference $h_{(n+1)} - h_{(n)}$ between consecutive shape functions $h_{(n)}$ for $n=1,2,3$ in embedding flow (black 3-2, blue 2-1, light blue 1-0).
  Right: Embedding flow at $t_{n=0,2,3}=0.0,0.66,1.0$ (outer-transparent, inner-transparent, inner-solid). 
\label{fig:emb_flow_contour}}
\end{figure}

\begin{figure}[h]
\includegraphics[width=0.48\linewidth]{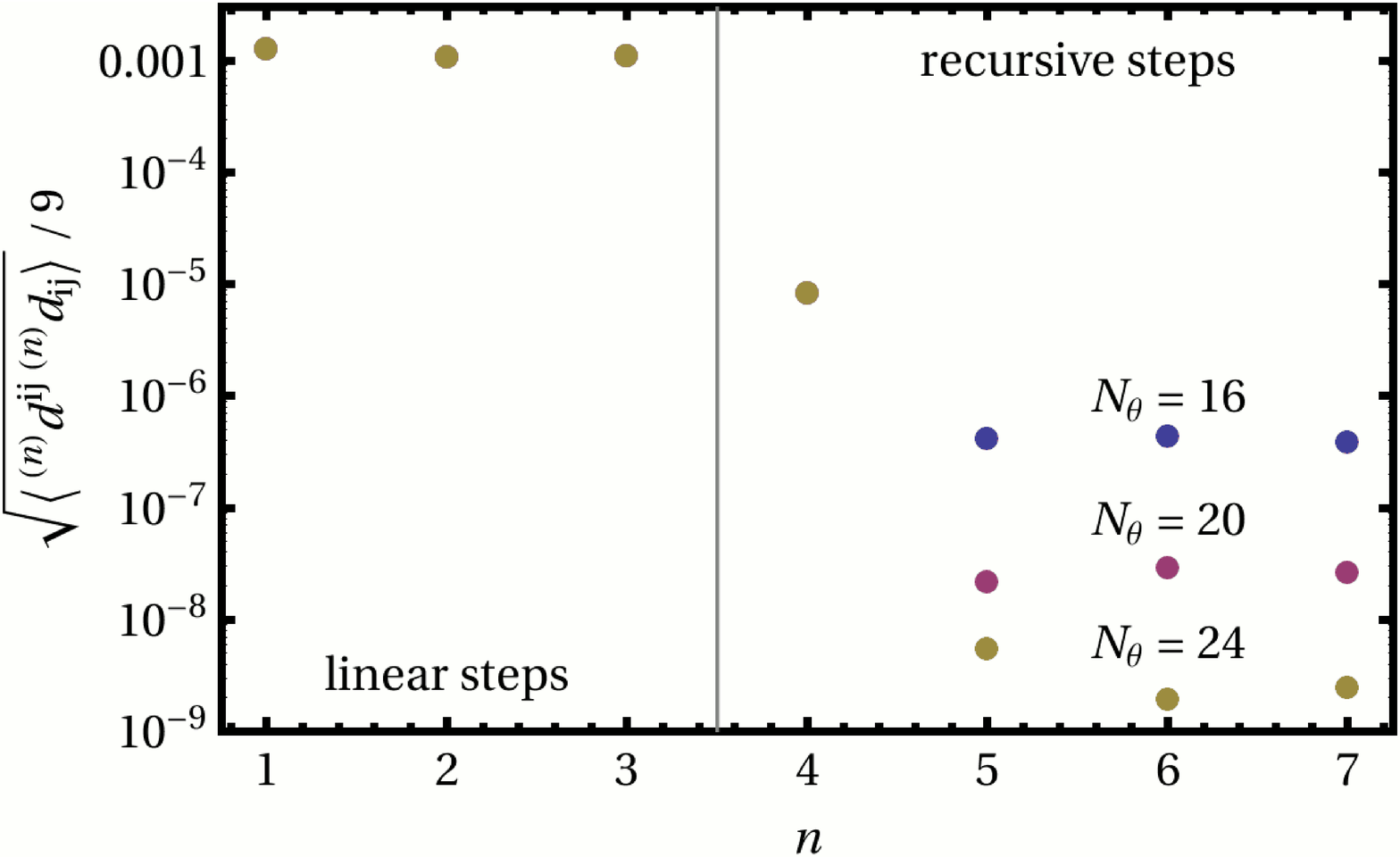} 
\includegraphics[width=0.48\linewidth]{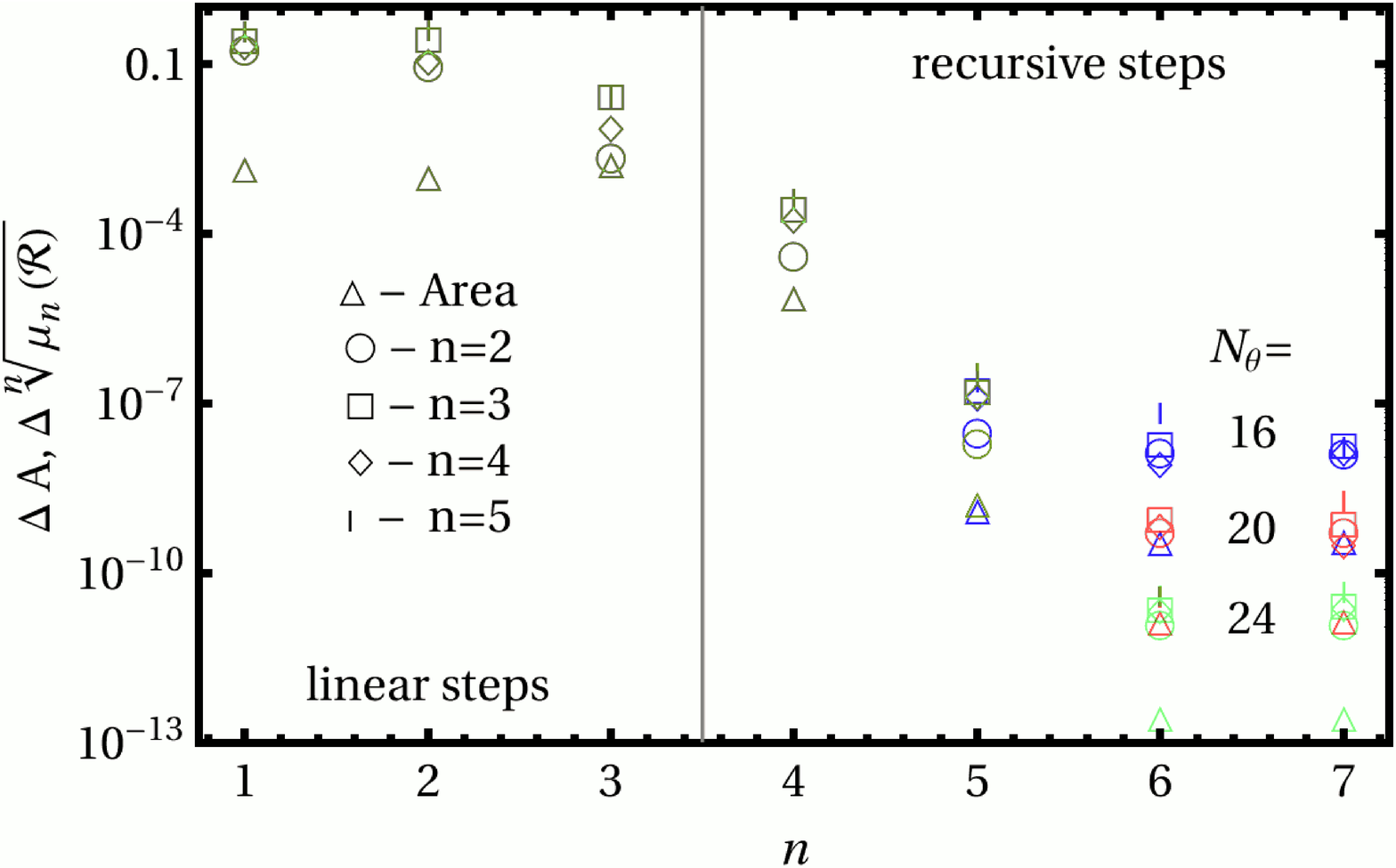} 
\caption{ Left: Difference between the flow and the induced metric at linear $t_{n=1,2,3}=0.33,0.66,1$ and recursive iteration $t_{n=4-7}=1$ of the embedding flow. 
  Right: Difference of the area $A$ and $\mu_n(\rsc)$ between the induced and target metric.
\label{fig:embedding_flow2}}
\end{figure}

\begin{figure}[h]
\includegraphics[width=0.48\linewidth]{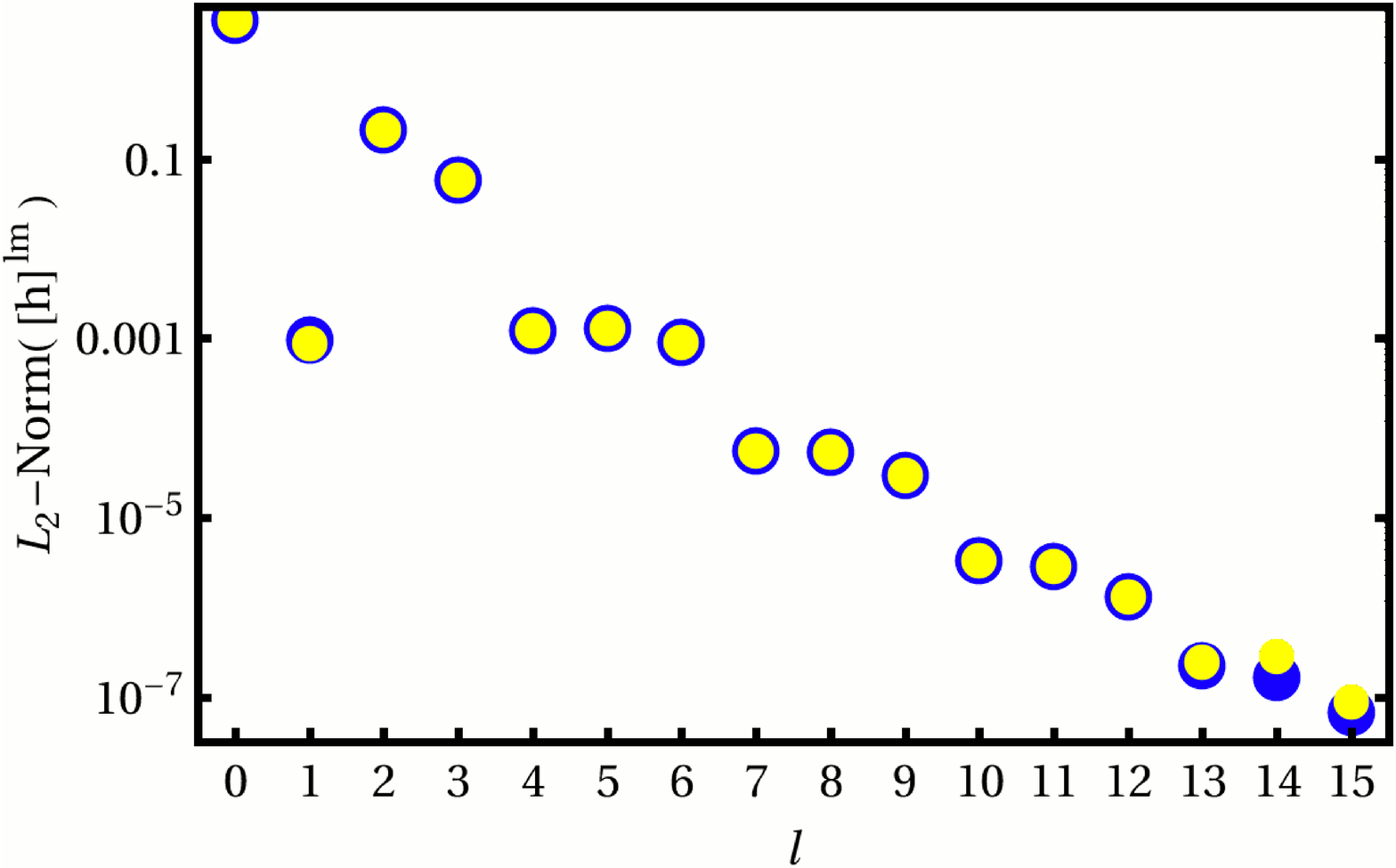}\quad 
\includegraphics[width=0.48\linewidth]{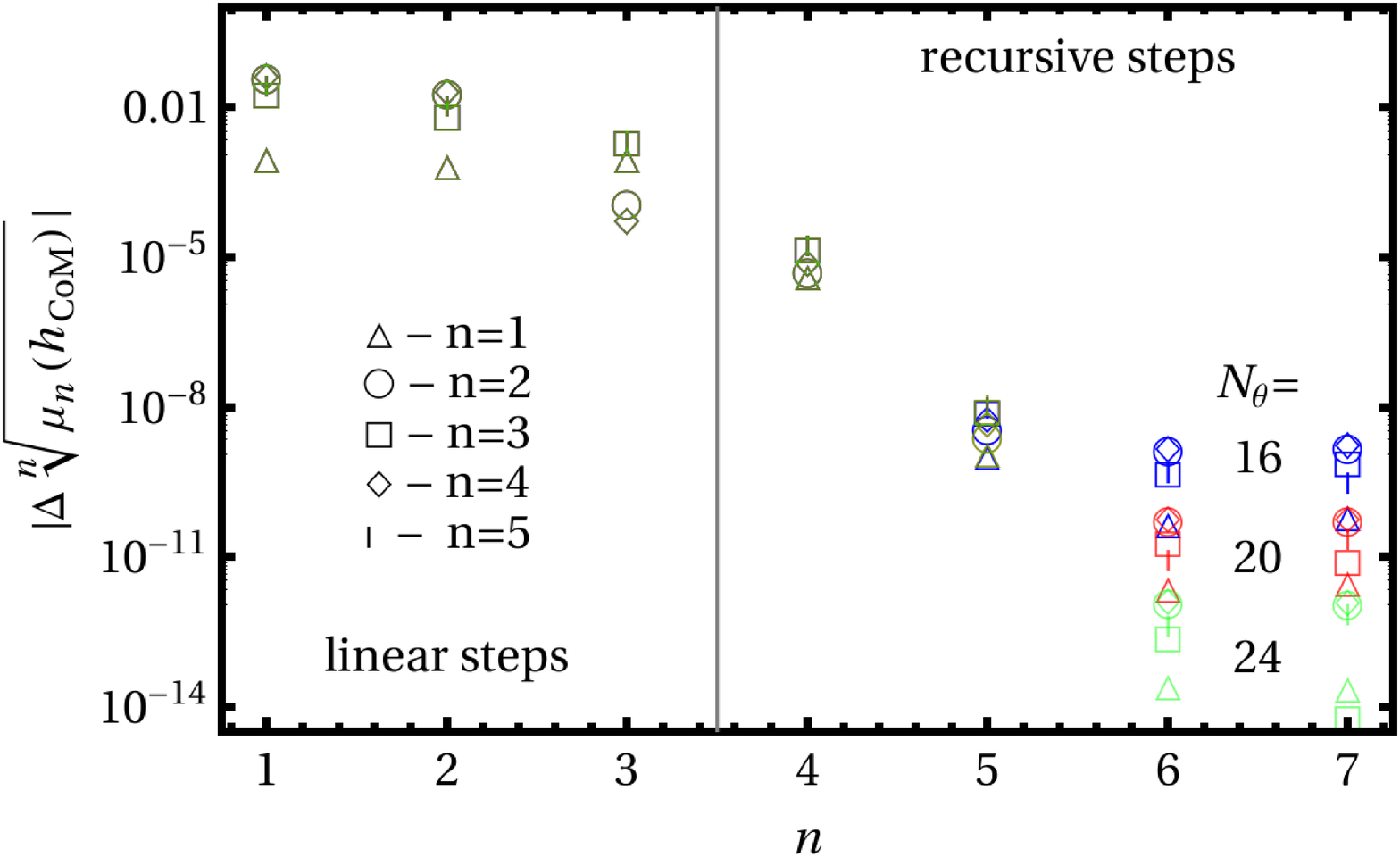} 
\caption{ Left: $L_2$-norm: $\sqrt{ \sum^{l}_{m=-l} ( [h]^{lm} )^2 }$ for original shape function $h$ and $h_{n=7}$ for $N_\theta=16$. The difference 
  for higher modes and $l=1$ is apparent.
  Right: Difference between the centered original and final shape function's $\mu_n$ converges with increasing resolution and flow time
   down to trunction error.
\label{fig:l2norms}} 
\end{figure}

We choose a conformal factor that is linear in $t$ then the change of $h_{(n)}$ during the first steps of
the flow $n=1,2,3$, $t_{n=0,1,2,3}=0.0,0.33,0.66,1.0$. It is about linear in $t$ as well which can be seen in
fig. (\ref{fig:emb_flow_contour}) (left); the difference between consecutive shape functions is approximately
constant along a fixed direction. The increase/decrease is strongest in the directions where the target metric
differs the most from the round metric. This is also apparent in fig. (\ref{fig:emb_flow_contour}) (left) and
fig. (\ref{fig:embedding_flow2}), where the embeddings of $q(t,\sigma)_{ij}$ at different states of the flow are
shown. The drifting of grid points as illustrated in fig. (\ref{fig:emb_flow}) is not visualised but it can be
seen how the unit sphere stepwise morphs into the target shape. The difference between $h_{(3)}$ and the original
shape function, see fig. (\ref{fig:embedding_flow2}), is about $10^{-2}$; in order to refine the solution further
we recursively apply the LEE, whereby $d_{ij}$ the difference between target metric and induced metric as well
as the intrinsic curvature converges exponentially with increasing $n$ and increasing resolution, limited by a
plateau of truncation error, see fig. (\ref{fig:l2norms}). Note that due to the non-uniqueness of the embedding
the final shape function is arbitrarily shifted and rotated wrt to the original one. In fig. (\ref{fig:l2norms})
(left) the $L_2$-norm in each $l$-eigenspace is shown. It is independent of rotations but not of the shifts,
which is clearly visible in the different $l=1$ modes.

\begin{figure}[h]
\includegraphics[width=0.49\linewidth]{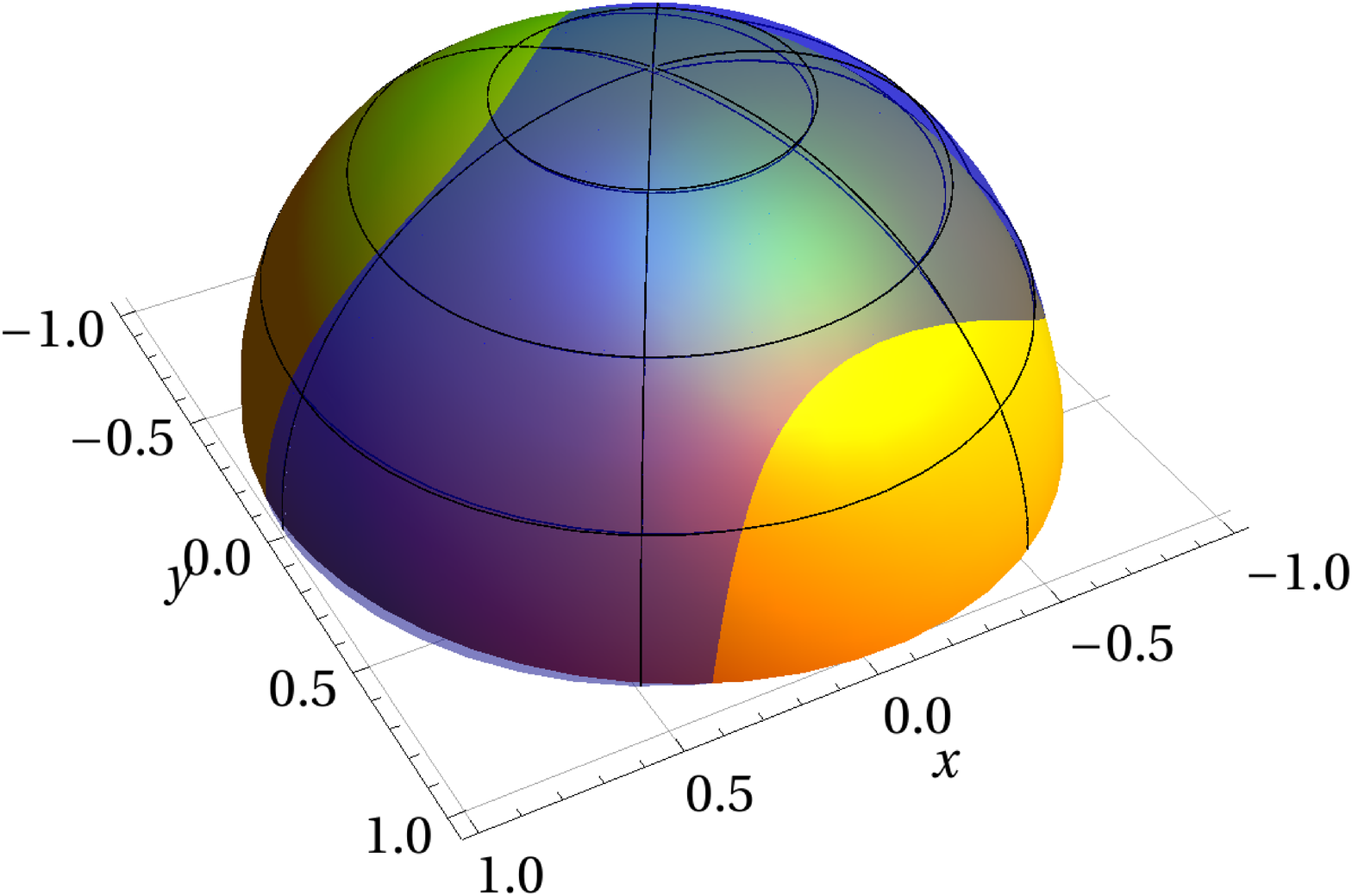} 
\includegraphics[width=0.49\linewidth]{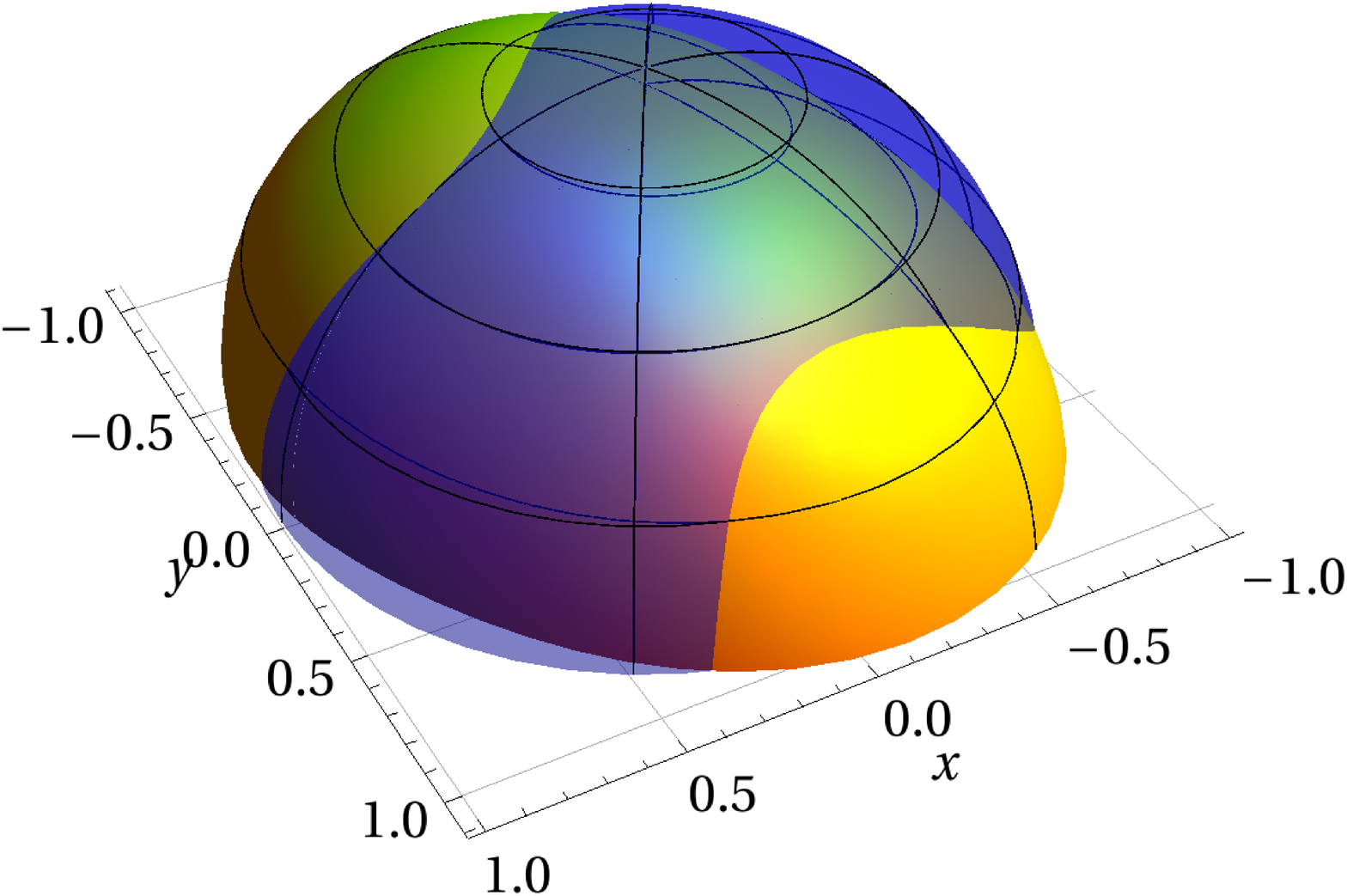} 
\caption{ Embedding flow around unit sphere (transparent) at $t_1=0.33$ (left) and $t_3=1$ (right)
\label{fig:embedding_flow}}
\end{figure}

\section{Conclusion}

In this article we presented a new numerical algorithm to solve the Weyl problem. The basic idea stems from the method
 of continuity which served as an approach in Nirenberg's proof of the conjecture and goes back to Weingarten and
Weyl. The linearized embedding equations play a central role. They can be reduced to a single linear elliptic PDE
by variable transformations. Solving the LEE stepwise for two 'nearby' metrics on a known embedding of one of the
metrics allows one to 'slowly' deform an initial shape into the desired embedding (embedding flow). In doing so,
it is necessary to link the target metric to an initial metric, whose embedding into $\RR^3$ is known. For this
purpose the round metric is well suited which is a steady-state solution of the Ricci flow on $S^2$ that can be
evolved from any metric. But it is then in general given in arbitrary coordinates.

Hence, apart from solving the embedding equations, a number of additional obstacles appear, which complicates a
numerical implementation of the method of continuity, namely we need to find a suitable coordinate system, solve
various (non)-linear elliptic PDEs,(anti)-differentiate functions on the surface, handle the drifting of grid
points under the embedding flow or the $SO(3,1)$ freedom of the solution etc. Another challenge is to keep the
total numerical error of all steps small.

Through the use of over-determined quasi-Cartesian coordinates on the surface as well as spectral methods and the
parabolic flow relaxation method we have shown how to overcome these technical difficulties altogether without
loosing numerical accuracy. Note that the compatibility with the simulations in $(3+1)$ numerical relativity is
automatically given, since our approach requires as input an arbitrary shape function as well as a Riemannian
3-manifold, both of which generate an admissible 2-metric on the $S^2$ surface. This allows direct applications
in binary black hole simulations to measure quasi-local mass or to investigate the large-scale dynamics of
numerical simulations of inhomogeneous cosmological models, where the coarse-graining process involves solving
the isometric embedding problem. These quasi-local mass definitions, like the Brown-York or the Kijowski-Liu-Yau
mass \cite{kijowski-1997-29,Brown:1991gb,Brown:1992br,Liu:2003bx}, are based on the comparison principle: anchor
the intrinsic geometries by isometric embeddings and compare the extrinsic geometries. Moreover, isometric
embeddings are ideal for visualising 2-metrics because they are fixed by unique shapes in the Euclidean space.

We have implemented our method with Fortran90 and tested it on a concrete arbitrarily chosen initial shape function
in the Euclidean space. By direct comparison between the final and initial shape functions and metrics we were able to demonstrate 
exponential convergence by increasing the spherical resolution as well as the number of recursive iteration steps of
the embedding flow. Our method is limited to positively curved and positive-definite 2-metrics and to ones 
whose isometric embedding and shape function in $\RR^3$ is 'well represented' through harmonic polynomials within
$\lmax<40$, i.e. whose spherical harmonics power spectrum drops quickly enough, but is then guaranteed
to reach a solution at reasonable computational costs, a few minutes on a modern CPU. 

Apart from the applications already mentioned the numerical methods in the sub-steps of our approach, like the
implementation to solve the Ricci-flow or the $l=1$ EV problem could lead to other applications in numerical
relativity on their own. For example, to solve other elliptic PDEs that appear to find approximate Killing vector
fields on trapping horizons \cite{Cook:2007wr} or to correct the gravitational waves signal of binary black hole
simulations that is commonly extracted on coordinate spheres at finite radius. In general the spherical coordinates
are distorted wrt the $l=1$ eigenfunctions of these spheres. Furthermore, there are quasi-local mass and momentum
measures, see \cite{Wang:2009} and references therein, which require isometric embeddings into Minkowski space
(and into $\RR^3$ as a sub-step) that would allow one to measure the linear momentum in numerical BBH simulation
quasi-locally.

MK would like to thank Lars Andersson and Luciano Rezzola for their invitation to the Max Planck Institute for
Gravitational Physics in Potsdam, which has enabled us to continue collaborating on this project. MJ thanks
Badri Krishnan, Alex Nielsen, Frank Ohme and Joachim Frieben for helpful comments and discussions. This work was
supported by the IMPRS for Gravitational Wave Astronomy in the Max Planck Society. 

\appendix
\section{Removing the positive eigenvalue from the spectrum of $\ML$}

In the code we solve equation (\ref{eqLw}) by relaxation method. This requires the linear operator $\ML$ 
to be non-negative. This is not the case because of the positive second term in (\ref{eqLdef}). 
$\ML$ a positive principal eigenvalue $\lambda_0$ but can be modified $\ML$ to push that
eigenvalue to the negative part of the spectrum. If the surface is not very far from a round sphere, we may
expect that the spectrum of $\ML$ will not be very different from the one of $\Delta + 2$, and in particular
will contain \emph{only one} positive eigenvalue. In fact, our numerical evidence supports the conjecture that
it is always the case for a convex surface.

The eigenfunction corresponding to $\lambda_0$ is not known, so the modification of $\ML$ is not completely trivial. 
First we define a new self--adjoint operator $\widetilde\ML$ as
\begin{eqnarray}
 \widetilde\ML(w) = \frac{1}{\sqrt{\MK}}\,\ML\left(\frac{w}{\sqrt{\MK}}\right), \label{eqTildeML}
\end{eqnarray}
with $\MK$ denoting the trace of $K_{AB}$ (we assume $K_{AB}$ is negative definite). Since $\sqrt{\MK}$ is positive
everywhere, the dimensionality of subspaces where $\WML$ is positive, negative or zero is the same as in $\ML$. In
particular, $\WML$ has only one positive eigenvalue, which is non--degenerate. One checks straightforward that
$\widetilde\ML\left(\sqrt{\MK}\right) = \sqrt{\MK}$, so this eigenvalue and the corresponding function are known
explicitly. We now modify $\ML$ further by defining
\begin{eqnarray}
 \HML(g) = \WML(g) - \frac{C}{\left\langle\sqrt{\MK},\sqrt{\MK}\right\rangle}\,\left\langle\sqrt{\MK},
g\right\rangle\,\sqrt{\MK}
\end{eqnarray}
for a constant $C>1$. It has the only positive eigenvalue shifted to the negative part of the spectrum and thus can be tackled 
by the relaxation method. Namely, the parabolic equation
\begin{eqnarray}
 \dot f = \HML(f) - \frac{\tau}{\sqrt{\MK}} + \frac{C}{\left\langle\sqrt{\MK},\sqrt{\MK}\right\rangle}\,
\left\langle\sqrt{\MK},\frac{\tau}{\sqrt{\MK}}\right\rangle\,\sqrt{\MK}, \label{eqdotf}
\end{eqnarray}
 converges to the solution of $(\ref{eqLw})$ divided by $\sqrt{\MK}$.
We can simplify (\ref{eqdotf}) further by introducing a new variable $u = \frac{f}{\sqrt{\MK}}$ which 
then (\ref{eqdotf}) becomes
\begin{eqnarray}
 \dot u = \frac{1}{\MK}\,\left(\ML(u) - \tau\right) - \frac{C}{\left\langle\sqrt{\MK},\sqrt{\MK}\right\rangle}\,
\left\langle 1,\ML(u) - \tau\right
 \rangle.
\end{eqnarray}
$u$ tends exponentially to a solution of (\ref{eqLw}). We have found out that this also holds true if we simplify this 
formula by replacing $C/\langle\sqrt{\MK},\sqrt{\MK}\rangle$ by a sufficiently large, positive number.

\section{Anti--differentiating a function in spherical harmonics}

In Section \ref{subslinemb} we show that the vector field $Y^i$ can be reconstructed from its derivatives on the
sphere. This leads us to the following numerical problem: Given the gradient of a function on $S^2$ expanded
in terms of spherical harmonics, what is the spherical harmonics decomposition of the original function?

Let $f$ be a function expanded in terms of $Y^{lm}$ 
\begin{eqnarray}
 f = \sum_{l=0}^{+\infty} \sum_{m=-l}^{m=l}\,^Y\!\!\left[f\right]^{lm}\,Y^{lm}.
\end{eqnarray}
The $\phi$ derivative is 
\begin{eqnarray}
 f_{,\phi} = \sum_{l=0}^{+\infty} \sum_{m=-l}^{m=l}\,^Y\!\!\left[f\right]^{lm}\,im\,Y^{lm},
\end{eqnarray}
so for all coefficients with $m\neq 0$ we obtain a straightforward relation between the expansions of
$f$ and $f_{,\phi}$:
\begin{eqnarray}
 \,^Y\!\!\left[f\right]^{lm} = -\frac{i}{m}\,^Y\!\!\left[f_{,\phi}\right]^{lm}.
\end{eqnarray}
In general the $\theta$ derivative of a regular $f$ has a 
discontinuity at $\theta=0,\pi$. This can be cured by multiplying it by $\sin\theta$. It is easy to see that
$\sin\theta\frac{\partial}{\partial\theta} Y^{l,m}$ can be expressed in terms of other spherical harmonics
with the same $m$. In particular, the harmonics with $m=0$ satisfy
\begin{eqnarray}
 \sin\theta \frac{\partial}{\partial \theta} Y^{l,0} = \frac{l(l+1)}{\sqrt{(2l+1)(2l+3)}}\,Y^{l+1,0} +
 \frac{l(l+1)}{\sqrt{(2l-1)(2l+1)}}\,Y^{l-1,0}.
\end{eqnarray}
This equation gives us a recurrence relation for the $m=0$ coefficients of $f$ in terms of 
$\sin\theta \frac{\partial f}{\partial\theta}$
\begin{eqnarray}
 \,^Y\!\!\left[f\right]^{l,0} =&& -\frac{\sqrt{(2l-1)(2l+1)}}{l(l+1)}\,^Y\!\!\left[\sin\theta f_{,\theta}\right]^{l-1,0}
+\nonumber \\
 && +\frac{(l-1)(l-2)}{l(l+1)}\,\sqrt{\frac{2l+1}{2l-3}}\,^Y\!\!\left[\sin\theta f_{,\theta}\right]^{l-2,0}
\end{eqnarray}
valid for $l \ge 2$. For $l=1$ the relation reads
\begin{eqnarray}
 \,^Y\!\!\left[f\right]^{1,0} = -\frac{\sqrt{3}}{2}\,^Y\!\!\left[\sin\theta f_{,\theta}\right]^{0,0}.
\end{eqnarray}
Finally the $\,^Y\!\!\left[f\right]^{0,0}$ coefficient is arbitrary. This corresponds to the possibility of adding a constant to the
solution.

\bibliographystyle{plain}
\bibliography{isom}

\end{document}